\def \s{~\rm{s}}
\def \km{~\rm{km}}

\def \yr{~\rm{yr}}
\def \pc{~\rm{pc}}
\def \kpc{~\rm{kpc}}
\documentclass[12pt,preprint]{aastex}
\usepackage{natbib}

\shortauthors{Soker}

\begin{document}

\title{CORRELATION OF BLACK HOLE-BULGE MASSES BY AGN JETS}

\author{Noam Soker\altaffilmark{1}}

\altaffiltext{1}{Dept. of Physics, Technion, Haifa 32000, Israel;
soker@physics.technion.ac.il.}

\begin{abstract}
I propose a feedback model to explain the correlation between the supermassive black hole (SMBH)
mass and the host galaxy bulge mass.
The feedback is based on narrow jets that are launched by the central SMBH, and
expel large amounts of mass to large distances. The condition is that the jets do not
penetrate through the inflowing gas, such that they can deposit their energy
in the inner region where the bulge is formed. For that to occur, the SMBH must
move relative to the inflowing gas, such that the jets continuously encounter fresh
gas. Taking into account the relative motion of the SMBH and the inflowing gas
I derive a relation between the mass accreted by the SMBH and the mass that is not
expelled, and is assumed to form the bulge. This relation is not linear, but rather
the SMBH to bulge mass ratio increases slowly with mass.
The same mechanism was applied to suppress star formation
in cooling flow clusters, making a tighter connection between the feedback
in galaxy formation and cooling flows.
\end{abstract}


\section{INTRODUCTION}
\label{sec:intro}

The tight correlation between the supermassive black hole (SMBH) mass, $M_{\rm BH}$, and the velocity
dispersion, $\sigma$, of the hot component of the host galaxy (e.g., Merritt \& Ferrarese 2001;
Gebhardt et al. 2000; G\"ultekin et al. 2009 ) is well established.
A recent paper by G\"ultekin et al. (2009 and earlier references therein)
derives the relation as
\begin{equation}
\log (M_{\rm BH}/M_\odot) =  (8.12 \pm 0.08) + (4.24\pm 0.41) \log (\sigma/200 \km \s^{-1}).
\label{eq:corr1}
\end{equation}
There is still no consensus on the slope, though. Shen et al. (2008), for example,
find the slope in luminous broad-line AGN to be $\sim 3.1-3.6$ instead of 4.24.

The correlation between the SMBH mass and the the bulge mass, $M_{\rm bulge}$
(e.g., Kormendy \& Richstone 1995; Laor 2001), is less tigh.
Using the $M_{\rm BH} \propto L^{1.11\pm0.18}$ correlation as derived by G\"ultekin et al. (2009),
where $L$ is the bulge stellar luminosity,  and substituting $L$ from the
$M_{\rm bulge}\propto L^{1.18}$ relation from Magorrian et al. (1998),
one derives $M_{\rm BH} \propto M_{\rm bulge}^{\beta}$ with $\beta=0.94 \pm 0.15$,
which is compatible with a linear relation $\beta=1$.
However, Laor (2001) derives a value of $\beta \simeq 1.2-1.8$, depending on the exact samples and
relations used.
Laor then deduced that there is no linear relation between $M_{\rm BH}$ and $M_{\rm bulge}$,
but rather that the ratio $M_{\rm BH}/M_{\rm bulge}$ increases from $\sim 0.05 \%$ in the least luminous
bulges ($M_V \sim -18$) to $0.5 \%$ in the bright ellipticals ($M_V \sim -22)$.

To further explore this relation, I examine the correlation
among the 20 brightest and 20 faintest galaxies in the sample of 31 galaxies used by
G\"ultekin et al. (2009) in their figure 4.
I find that for the brightest galaxies the value of $\gamma$ in the
relation $M_{\rm BH} \propto L^{\gamma}$ is larger by $\sim 25 \%$ than the value of
$\gamma$ for the entire sample, while for the 20 faintest galaxies it is lower by
$\sim 14 \%$.
Using then the relation $M_{\rm bulge}\propto L^{1.18}$ (Magorrian et al. 1998), one finds
that the ratio $M_{\rm BH}/M_{\rm bulge}$ increases with mass, as claimed by Laor (2001),
and obtained in some models and simulations (e.g., Shabala \& Alexander 2009).

The relation $L -\sigma$ is not tightly constrained either.
Laor (2001), citing Nelson \& Whittle (1996), takes $L \sim \sigma^{3}$,
and using $M_{\rm bulge}\propto L^{1.18}$ (Magorrian et al. 1998) gets
$M_{\rm bulge}\propto \sim \sigma^{3.5}$.
Using this relation in equation (\ref{eq:corr1}) gives $\beta=1.2$.
H\"aring \& Rix (2004) find $\beta=1.12\pm 0.06$, and
$M_{\rm BH}/M_{\rm bulge}=0.14\% \pm 0.04\%$ for $M_{\rm bulge} \sim 5\times 10^{10} M_\odot$.

Based on these considerations, and for the purpose of the present paper, I assume
the deviation from a linear $M_{\rm BH} - M_{\rm bulge}$ relation to be due to an additional
dependance on the dispersion velocity
\begin{equation}
M_{\rm BH} = \Theta  M_{\rm bulge} \left ( \frac{\sigma}{200 \km \s^{-1}} \right)^\chi,
\label{eq:corr2}
\end{equation}
where $\Theta \simeq 10^{-3}$ and $\chi \simeq 0-1$.
The above expression is an average one, as the relations between the SMBH mass, the bulge mass,
and the dispersion velocity are somewhat different for elliptical galaxies, classical
bulges and pseudo-bulges (Gadotti \& Kauffmann 2009).
In this first presentation of the idea I ignore these differences.

In this paper I will try to account for relation (\ref{eq:corr2}) with a feedback mechanism
based on jets launched by the SMBH.
The feedback mechanism where AGN jets (outflow; wind) suppress gas from cooling to low temperatures and from
forming stars was discussed for both cooling flows in galaxies and clusters of galaxies
(e.g., Binney \& Tabor 1995; Nulsen \& Fabian 2000; Reynolds et al. 2002;
Omma \& Binney 2004; Soker \& Pizzolato 2005),
and in galaxy formation (e.g., Silk \& Rees 1998; Fabian 1999; King 2003;
Croton et al. 2006; Bower et al. 2008; Shabala \& Alexander 2009).
I note that some of the papers (e.g., Silk \& Rees 1998; King 2003) make use of the
Eddington luminosity limit; the model proposed here make no use of the Eddington luminosity limit.
Most models (e.g., Silk \& Rees 1998; Fabian 1999) do not consider the geometry explicitly; here
the geometry of the narrow jets and the motion of their source are key issues.
Bower et al. (2008) numerically tried to derive the $M_{\rm BH}-M_{\rm bulge}$ correlation.
In the present paper I try to present the basic physics that I propose leads to this
correlation.

\section{THE JET-FEEDBACK MECHANISM}
\label{sec:jet}

I do not consider the formation of the seed SMBH, but rather assume its existence.
The proposed mechanism is based on the following assumptions.
\begin{enumerate}
\item The feedback mechanism, i.e., the one that expels mass from the inner regions
to large distances and heats it up before stars can be formed, is driven by jets launched
by an accretion disk around the SMBH. This is supported by the presence of X-ray deficient bubbles
in clusters of galaxies that show that AGN jets are sufficiently energetic to heat the
intracluster medium (see review by McNamara \& Nulsen 2007).
\item The properties of jets launched by SMBH have some universal properties,
such as the fast jet's speed $v_f \simeq c$.
\item There is a universal efficiency (e.g., Tremaine 2005) of converting accretion energy to
kinetic energy of the two jets.
Considering different works (e.g., K\"ording et al. 2008; Merloni \& Heinz 2008) and
taking into account the result of Binney et al. (2007), I take this ratio to be $\sim 0.05$.
As the jets are launched at velocities of $v_f \simeq c$, I take the ratio of mass
launched in the two jets to accreted mass to be $\eta \equiv \dot M_f/\dot M_{\rm acc} \simeq 0.05$.
\item The mass flowing in at early stages, i.e., the mass available for star formation,
is very large. Namely, the mass that is converted to stars is limited by the feedback mechanism and
not by the mass available in the SMBH surroundings. This is supported by studies of galaxy formation
(e.g., Bower et al. 2008).
\item There is a relative motion between the SMBH and the inflowing mass of
$v_{\rm rel} \simeq \sigma$, as the relaxation time of the SMBH is longer than the
galaxy formation time scale (Tremaine 2005).
\item The surrounding mass $M_s$ that resides at a typical distance $r_s$ and
having a density $\rho_s$, (see below) is
flowing inward at a velocity of $\sim \sigma$.
Thus, $\dot M_s \simeq 4 \pi r_s^2 \sigma \rho_s$,
and it is resupplied on a time scale of $\sim r_s/\sigma$.
This assumption should be better constrained by future 3D numerical simulations.
\end{enumerate}

With these assumptions in hand, I proceed to describe the interaction of the jets with
the surrounding inflowing gas.
If the jets penetrate through the surrounding gas they will be collimated by that gas,
and two narrow collimated fast jets will be formed, similar to the flow structure in
the simulations of Sutherland \& Bicknell (2007).
By fast it is understood that the jet's velocity is not much below its original velocity.
If, on the other hand, the jets cannot penetrate the surrounding gas they will
accelerate the surrounding gas and form SMW (slow-massive-wide) outflow (Soker 2008).
The conditions for the jets not to penetrate the surrounding gas but rather form a SMW outflow
are derived below.
The derivation will not include relativistic effects, although the fast jets might be relativistic.
The other uncertainties involved are larger than those introduced by this assumption
(assuming the jets are not highly relativistic).

Let the jets from the inner disk zone have a mass outflow rate in both directions
of $\dot M_f$, a velocity $v_f$, and let the two jets cover a solid angle of
$4 \pi \delta$ (on both sides of the disk together).
The density of the outflow at radius $r$ is
\begin{equation}
\rho_f =  \frac {\dot M_f}{4 \pi \delta r^2 v_f}.
\label{eq:rhof}
\end{equation}
Let the jets encounter the surrounding gas residing within a distance
$r_s$ and having a typical density $\rho_s$.
The head of each jet proceeds at a speed $v_h$ given by the balance
of pressures on its two sides.
Assuming supersonic motion this equality reads
$\rho_s v_h^2 = \rho_f (v_f-v_h)^2$, which can be solved for $v_h$
\begin{eqnarray}
\frac {v_f}{v_h}-1 =
\left( \frac {4 \pi \delta r_s^2 v_f \rho_s}{\dot M_f} \right)^{1/2}
\simeq \left( \frac {\delta \dot M_s v_f }{\dot M_f \sigma} \right)^{1/2}
\nonumber \\
=1225
\left( \frac {\dot M_s/\dot M_f}{10^4} \right)^{1/2}
\left( \frac {\delta}{0.1} \right)^{1/2}
\left( \frac {v_f}{c} \right)^{1/2}
\left( \frac {\sigma}{200 \km \s^{-1}} \right)^{-1/2}.
\label{eq:vh1}
\end{eqnarray}
where in the second equality the mass inflow rate $\dot M_s \simeq 4 \pi \rho_s \sigma r_s^2$
(by assumption 6), has been substituted.
The time required for the jets to cross the surrounding gas and break out of it is given by
\begin{equation}
t_p \simeq \frac {r_s}{v_h} \simeq
\frac{r_s}{v_f}
\left( \frac {\delta \dot M_s v_f }{\dot M_f \sigma} \right)^{1/2}
=4 \times 10^6
\left( \frac {r_s}{1 \kpc} \right) \yr ,
\label{eq:tp1}
\end{equation}
where in the last equality the same values as in equation (\ref{eq:vh1}) have been used.

If there are no changes in the relative geometry of the SMBH and inflowing mass,
the jets will rapidly penetrate the surrounding gas and expand to large distances.
In this case the jets will not deposit their energy in the inflowing gas.
For an efficient deposition of energy to the inflowing gas, we require that there will
be a relative motion between the SMBH and the inflowing gas, such that the jets
continuously encounter fresh mass.
The relevant time is the time that the transverse motion of the jet
crosses it width $\tau_s \equiv D_j/v_{\rm rel} \simeq D_j/\sigma$,
as by our assumption 5 the relative velocity is $v_{\rm rel} \simeq \sigma$.
The width of the jet at a distance $r_s$ from its source is $D_j=2 r_s \sin \alpha$,
where $\alpha$ is the half opening angle of the jet. For a narrow
jet  $\sin \alpha \simeq \alpha \simeq (2 \delta)^{1/2}$, and
\begin{equation}
\tau_s = \frac {2 (2 \delta)^{1/2} r_s}{\sigma}.
\label{eq:taus}
\end{equation}

The demand for efficient energy deposition, $\tau_s \la t_p$, reads then
\begin{equation}
\frac {\dot M_s}{\dot M_f} \ga
8   \frac {v_f}{\sigma}.
\label{eq:ms1}
\end{equation}
{{{ This result can be understood as follows. The ratio ${v_f}/{\sigma}$ comes from the ratio
of the ram pressure of the narrow jet to that of the ambient gas which disturbs the jet, and from
the relative transverse motion of the jet and the abient gas.
The number 8 comes from the geometry of a narrow jet with a relative transverse velocity to
that of the ambient gas. }}}
Using assumption 3 that $\dot M_f=\eta \dot M_{\rm acc}$, and substituting typical values, this condition reads
\begin{equation}
\frac {\dot M_s}{\dot M_{\rm acc}} \ga
8 \eta
\frac {v_f}{\sigma} = 600
\left( \frac {\eta}{0.05} \right)
\left( \frac {v_f}{c} \right)
\left( \frac {\sigma}{200 \km \s^{-1}} \right)^{-1} .
\label{eq:mscluster}
\end{equation}

The accretion rate $\dot M_{\rm acc}$ is the accretion rate onto the SMBH, and the
inflow rate of the surrounding gas is assumed to form stars in the bulge (if it is not
expelled by the jets).
If the inflow rate is above the value given in equation (\ref{eq:mscluster}), the
deposition of energy by the jets is efficient enough to expel the mass back to large
distances and heat it (Soker 2008).
Following assumption 4 above, I take the SMBH to bulge mass ratio
to be equal to $\dot M_{\rm acc}/{\dot M_s}$.
Equation (\ref{eq:mscluster}) yields
\begin{equation}
M_{\rm BH}= 0.0017 M_{\rm bulge}
\left( \frac {\eta}{0.05} \right)^{-1}
\left( \frac {v_f}{c} \right)^{-1}
\left( \frac {\sigma}{200 \km \s^{-1}} \right) .
\label{eq:mBH2}
\end{equation}

The last equation has the form of equation (\ref{eq:corr2}), and by using typical
properties of AGN jets and bulges ($\eta$; $v_f$; $\sigma$), the
numerical value of equation (\ref{eq:corr2}) is reproduced with $\chi=1$.
{{{ Basically, the equation contains one parameter $p_m \equiv \eta v_f$.
Its physical meaning is that of the momentum of the material ejected in the jets per
unit accreted mass to the SMBH.  }}}
For a more detailed comparison to observations, future work is needed to incorporate the
proposed feedback mechanism into numerical simulations of galaxy formation
and evolution, e.g., such as those described by Di Matteo et al. (2005).

\section{DISCUSSION AND SUMMARY}
\label{sec:summary}

I built a simple feedback mechanism where jets launched by the central supermassive black hole (SMBH)
interact with the surrounding inflowing gas in a young galaxy.
By using several very plausible assumptions, most of which are well supported by
observations, I derived a relation between the mass that is converted to stars and the
mass accreted by the central SMBH (eqs. \ref{eq:mscluster} and \ref{eq:mBH2}).
The basic process to suppress star formation requires that the jets do not
penetrate the surrounding gas; if they do, their energy is deposited at large
distances, rather than in the gas that potentially can form stars in the bulge.
For that, the jets should encounter new material before they escape. Namely,
the typical time for the jets to cross its own width at radius $r_s$ should be shorter
than the penetration time at radius $r_s$, $\tau_s \la t_p$.
This leads to equation (\ref{eq:mscluster}).

If the mass inflowing rate $\dot M_s$ is larger than the value given by
equation (\ref{eq:mscluster}) the jets are very efficient in depositing their energy to the
inflowing gas. The jets are shocked, and since the radiative cooling time is long
(Soker 2008), there is no energy loss and the jets can expel huge amounts of mass
back to large distances and heat the gas. This suppresses star formation.
Even the Compton cooling time (e.g., King 2003) is longer than the jet flow time
$r_s/v_f$ for $r_s \ga 10 \pc$, and here $r_s \sim 0.1-1 \kpc$.
I therefore take the mass inflow rate into the bulge to be as the limiting value
in equation (\ref{eq:mscluster}). This leads to equation (\ref{eq:mBH2}), which
has the form and numerical value as in equation (\ref{eq:corr2}).
{{{ In addition to the assumptions of the model, equation (\ref{eq:mBH2})
contains only one parameter $p_m \equiv \eta v_f$, which is the momentum of
the material ejected in the jets per unit accreted mass to the SMBH.  }}}

I note that the criterion that the jets do not penetrate the surrounding material
might be relevant also to models that attribute the correlation between the SMBH mass and
the host galaxy bulge mass to a negative feedback (e.g., Silk \& Norman 2009);
in a negative feedback model the jets induce star formation in the surrounding gas.

King (2003) compares the cooling time to the slow outflow time. Since the outflow gas has
a velocity $\la \sigma$ in his model, he finds cooling to be very important.
King assumes the presence of an almost spherical shell, and does not consider
its formation by two narrow jets as the present model does.
Once the condition found here is met, i.e., the jets do not penetrate the surrounding gas,
a slow expanding shell is formed, and some of the calculations of Silk \& Rees (1998),
Fabian (1999), and King (2003) are applicable here as well.
However, here I made no use of the Eddington luminosity limit, and a similar
mechanism can be applied to cooling flows (see below).

As the system evolves it relax. The relative motion of the SMBH and the inflowing gas decreases
with time. Therefore, as the system ages, and masses and $\sigma$ increase, condition
(\ref{eq:mscluster}) is more difficult to fulfill, and the accretion rate to the bulge can
be larger than given by equation (\ref{eq:mscluster}). Namely, the SMBH mass is lower than
given by equation (\ref{eq:mBH2}). This implies that the power of $\sigma$ there
is $\chi<1$, as is deduced from observations in equation (\ref{eq:corr2}).

Although the proposed feedback mechanism has several assumptions, they are all
very plausible. The feedback mechanism will have to be studied by 3D numerical
simulations, as the relative motion of the SMBH and the inflowing gas must be included.
There is a big advantage to the proposed mechanism, in that the same basic mechanism
can work to suppress star formation in cooling flows (Soker 2008). Using the
same mechanism with similar parameters, I developed a mechanism for the formation of
slow massive wide (SMW) outflows (or SMW jets) in cooling flow clusters (Soker 2008).
Such jets can inflate the observed X-ray deficient bubbles that are observed in
clusters of galaxies (Sternberg et al. 2007).

\acknowledgements
I thank Craig Sarazin for very helpful comments.
This research was supported by the Asher Fund for Space
Research at the Technion, and the Israel Science foundation.

\end{document}